\documentclass[english,runningheads,a4paper]{llncs}[2018/03/10]

\usepackage[ngerman,main=english]{babel}
\addto\extrasenglish{\languageshorthands{ngerman}\useshorthands{"}}
\usepackage{regexpatch}
\makeatletter
\edef\switcht@albion{%
  \relax\unexpanded\expandafter{\switcht@albion}%
}
\xpatchcmd*{\switcht@albion}{ \def}{\def}{}{}
\xpatchcmd{\switcht@albion}{\relax}{}{}{}
\edef\switcht@deutsch{%
  \relax\unexpanded\expandafter{\switcht@deutsch}%
}
\xpatchcmd*{\switcht@deutsch}{ \def}{\def}{}{}
\xpatchcmd{\switcht@deutsch}{\relax}{}{}{}
\edef\switcht@francais{%
  \relax\unexpanded\expandafter{\switcht@francais}%
}
\xpatchcmd*{\switcht@francais}{ \def}{\def}{}{}
\xpatchcmd{\switcht@francais}{\relax}{}{}{}
\makeatother

\usepackage{ifluatex}
\ifluatex
  \usepackage{fontspec}
  \usepackage[english]{selnolig}
\fi

\iftrue % use default-font
  \ifluatex
  \else
    \usepackage[%
      rm={oldstyle=false,proportional=true},%
      sf={oldstyle=false,proportional=true},%
      tt={oldstyle=false,proportional=true,variable=false},%
      qt=false%
    ]{cfr-lm}
  \fi
\else
  \ifluatex
  \else
    \usepackage{newtxtext}
    \usepackage{newtxmath}
    \usepackage[zerostyle=b,scaled=.9]{newtxtt}
  \fi
\fi

\ifluatex
\else
  \usepackage[T1]{fontenc}
  \usepackage[utf8]{inputenc} %support umlauts in the input
\fi

\usepackage{graphicx}

\usepackage{upquote}

\usepackage{booktabs}

\usepackage{paralist}

\usepackage{csquotes}

\usepackage{textcmds}

\RequirePackage[%
  babel,%
  final,%
  expansion=alltext,%
  protrusion=alltext-nott]{microtype}%
\DisableLigatures{encoding = T1, family = tt* }

\usepackage{url}
\makeatletter
\g@addto@macro{\UrlBreaks}{\UrlOrds}
\makeatother
\usepackage{xcolor}

\usepackage{listings}
\renewcommand{\lstlistingname}{Listing.}
\lstdefinelanguage{rebeca}{
  morekeywords={reactiveclass, knownrebecs, statevars, main, msgsrv, main, define, LTL, CTL, boolean, int, shortint, byte, if, else, while, for, wait, msg, reset, set, self, false, true, now, after, delay, deadline, initial, env,extends, abstract},
  otherkeywords={=>,<-,<\%,<:,>:,\#,@},
  sensitive=true,
  morecomment=[l]{//},
  morecomment=[n]{/*}{*/},
  morestring=[b]",
  morestring=[b]',
  morestring=[b]"""
}

\usepackage{chngcntr}
\AtBeginDocument{\counterwithout{lstlisting}{section}}

\usepackage{pdfcomment}
\usepackage[%
  square,        % for square brackets
  comma,         % use commas as separators
  numbers,       % for numerical citations;
  sort&compress, % as sort but in addition multiple numerical citations
]{natbib}
\ifluatex
\else
  \SetExpansion
  [ context = sloppy,
    stretch = 30,
    shrink = 60,
    step = 5 ]
  { encoding = {OT1,T1,TS1} }
  { }
\fi

\usepackage{stfloats}
\fnbelowfloat

\usepackage{hyperref}
\hypersetup{hidelinks,
  colorlinks=true,
  allcolors=black,
  pdfstartview=Fit,
  breaklinks=true}
\usepackage[all]{hypcap}

\usepackage[group-four-digits,per-mode=fraction]{siunitx}

\usepackage[capitalise,nameinlink]{cleveref}
\usepackage{iflang}
\IfLanguageName{ngerman}{
  \crefname{table}{Tab.}{Tab.}
  \Crefname{table}{Tabelle}{Tabellen}
  \crefname{figure}{\figurename}{\figurename}
  \Crefname{figure}{Abbildungen}{Abbildungen}
  \crefname{equation}{Gleichung}{Gleichungen}
  \Crefname{equation}{Gleichung}{Gleichungen}
  \crefname{listing}{\lstlistingname}{\lstlistingname}
  \Crefname{listing}{Listing}{Listings}
  \crefname{section}{Abschnitt}{Abschnitte}
  \Crefname{section}{Abschnitt}{Abschnitte}
  \crefname{paragraph}{Abschnitt}{Abschnitte}
  \Crefname{paragraph}{Abschnitt}{Abschnitte}
  \crefname{subparagraph}{Abschnitt}{Abschnitte}
  \Crefname{subparagraph}{Abschnitt}{Abschnitte}
}{
  \crefname{section}{Sect.}{Sect.}
  \Crefname{section}{Section}{Sections}
  \crefname{listing}{\lstlistingname}{\lstlistingname}
  \Crefname{listing}{Listing}{Listings}
}

\usepackage{xspace}
\DeclareFontFamily{U}{MnSymbolC}{}
\DeclareSymbolFont{MnSyC}{U}{MnSymbolC}{m}{n}
\DeclareFontShape{U}{MnSymbolC}{m}{n}{
  <-6>    MnSymbolC5
  <6-7>   MnSymbolC6
  <7-8>   MnSymbolC7
  <8-9>   MnSymbolC8
  <9-10>  MnSymbolC9
  <10-12> MnSymbolC10
  <12->   MnSymbolC12%
}{}
\DeclareMathSymbol{\powerset}{\mathord}{MnSyC}{180}

\ifluatex
\else
  \input glyphtounicode
\fi

\usepackage[math]{blindtext}
\usepackage{mwe}
\usepackage{comment}
\usepackage{tikz}
\usepackage{subfigure}
\usetikzlibrary{backgrounds,automata}
\usepackage{graphicx}
\usepackage{pgfplots}

\begin{document}

\newcommand{\FeriVANca}{VeriVANca\xspace}
\title{\FeriVANca: An Actor-Based Framework for Formal Verification of Warning Message Dissemination Schemes in VANETs}
%If Title is too long, use \titlerunning
\titlerunning{\FeriVANca}

%Single insitute
\author{Farnaz Yousefi\inst{1} \and Ehsan Khamespanah\inst{2,3} \and Mohammed Gharib\inst{4} \and Marjan Sirjani\inst{3,5} \and Ali Movaghar\inst{1}}
%If there are too many authors, use \authorrunning
\authorrunning{Farnaz Yousefi et al.}
\institute{
Department of Computer Engineering,
Sharif University of Technology - Iran
\and
School of Electrical and Computer Engineering,
University of Tehran - Iran
\and
School of Computer Science,
Reykjavik University - Iceland
\and
School of Computer Science,
Institute for Research in Fundamental Science - Iran
\and
School of IDT, 
M\"{a}lardalen University - Sweden
}

%% Multiple insitutes - ALTERNATIVE to the above
% \author{%
%     Firstname Lastname\inst{1} \and
%     Firstname Lastname\inst{2}
% }
%
%If there are too many authors, use \authorrunning
%  \authorrunning{First Author et al.}
%
%  \institute{
%      Insitute 1\\
%      \email{...}\and
%      Insitute 2\\
%      \email{...}
%}

\maketitle

\begin{abstract}
One of the applications of vehicular ad-hoc networks is warning message dissemination among vehicles in dangerous situations to prevent more damage. The only communication mechanism for message dissemination is multi-hop broadcast; in which, forwarding a received message have to be regulated using a scheme regarding the selection of forwarding nodes. When analyzing these schemes, simulation-based frameworks fail to provide guaranteed analysis results due to the high level of concurrency in this application. Therefore, there is a need to use model checking approaches for achieving reliable results. In this paper, we have developed a framework called \FeriVANca, to provide model checking facilities for the analysis of warning message dissemination schemes in VANETs. To this end, an actor-based modeling language, Rebeca, is used which is equipped with a variety of model checking engines. To illustrate the applicability of \FeriVANca, modeling and analysis of two warning message dissemination schemes are presented. Some scenarios for these schemes are presented to show that concurrent behaviors of the system components may cause uncertainty in both behavior and performance which may not be detected by simulation-based techniques. Furthermore, the scalability of \FeriVANca is examined by analyzing a middle-sized model.

\end{abstract}


\section{Introduction}\label{sec:intro}
%%%% references and section referencing
Vehicular Ad-hoc NETworks (VANETs) have attracted much attention in both academia and industry during the last years. The emergence of autonomous vehicles and the safety concerns regarding the use of these vehicles in the near future have highlighted the possible use of VANETs in safety enhancement of future transportation system. Using VANETs in such mission critical applications, calls for reliability assurance of algorithms. One of the applications in this domain is the use of vehicle to vehicle communication for Warning Message Dissemination (WMD) in dangerous situations to prevent further damage. In this application, vehicles broadcast warning messages to inform each other of the upcoming hazard. To increase the number of vehicles receiving the warning message, the receiving nodes should forward the message. To hold the trade-off between the traffic in the network and maximum number of vehicles receiving the message, a number of schemes regarding the selection of forwarding nodes have been proposed~\cite{DBLP:journals/mis/SanguesaFGMCC16}. More details about WMD in VANETs are presented in Section~\ref{sec:problemDef}.
 
A number of simulation-based tools and techniques have been used for the analysis of these WMD schemes. However, concurrent execution of system components reduces the effectiveness of simulation-based approaches for such mission critical applications. This is because of the fact that simulation-based approaches cannot provide high level of confidence for the correct behavior of the system. In such cases, there is a need to apply formal verification for achieving reliable results. Formal verification is used in applications of VANETs such as cooperative collision avoidance~\cite{DBLP:journals/tits/HafnerCCV13}, intersection management using mutual exclusion algorithms~\cite{DBLP:conf/icpp/AoxueluoWCR13}, and collaborative driving~\cite{DBLP:journals/jits/LinM16}. However, to the best of our knowledge, there is no work on formal verification of WMD application in VANETs. 

In this paper, we introduce \FeriVANca as a framework for the analysis of WMD schemes in VANETs. To this end, we develop \FeriVANca in Timed Rebeca~\cite{timed-rebeca}, a real-time extension of Rebeca~\cite{DBLP:journals/fuin/SirjaniMSB04}. Rebeca is an operational interpretation of the actor model with formal semantics, supported by a variety of analysis tools~\cite{DBLP:journals/scp/KhamespanahKS18}. In the actor model, all the elements that are running concurrently in a distributed system are modeled as actors. Communication among actors takes place by asynchronous message passing. These structures and features match the needs of VANETs as they consist of autonomous nodes which communicate by message passing. This level of faithfulness helps in having a more natural mapping between the actor model and VANETs, making models easier to develop and understand. In Section~\ref{sec:rebeca} Timed Rebeca is briefly introduced using the counting-based scheme example.

To illustrate the applicability of this approach, we have modeled a distance-based scheme~\cite{TLO} and a counting-based scheme~\cite{counting-based-scheme} using \FeriVANca. Results of model checking for the distance-based scheme show that concurrent execution of the system components enables multiple execution traces some of which cause starvation and may not be detected using simulation-based techniques (Section~\ref{sec:starvation}). 
We also observed that, in a given scenario, multiple numbers may be achieved for the performance when considering the interleaving of concurrently executing components. Our further investigations yield that having multiple performance results is not limited to one scenario and commonly happens. More details on these cases are presented in Section~\ref{nondet-performance-result}.
Furthermore, to examine the scalability of \FeriVANca, a middle-sized model of a four-lane street with about 40 vehicles is analyzed. We observed that if scaling up the number of vehicles results in creation of very congested areas, the size of the state space and analysis time is increased dramatically. However, scaling up the model without creation of new congested areas, results in smooth increase in the size of the state space and analysis time as presented in Section~\ref{sec:scalability}.

\section{Warning Message Dissemination in VANETs}
\label{sec:problemDef}

warning message dissemination (WMD) is an application developed for VANETs that tends to increase the safety and comfort of passengers.  In this application, a warning message is disseminated between vehicles in the case of any abnormal situations such as car accident or undesirable road conditions. Received warning messages are used either to activate an automatic operation or are shown as alerts to inform the driver of the upcoming hazard.

Using WMD in safety-critical applications, requires providing high reliability for the application in developed solutions. Besides, some characteristics of VANETs such as high mobility of the nodes and fast topology changes, makes routing algorithms commonly used in MANETs inapplicable to VANETs~\cite{DBLP:journals/telsys/ZeadallyHCIH12}. Therefore, the only approach for implementation of message dissemination in VANETs is multi-hop broadcast of the message. In this approach, the receiving nodes are responsible for re-broadcasting the message to the others. However, this can result in broadcast storm problem in the network. In order to tackle this problem, a number of schemes have been proposed for WMD as described in the following section. 
 
\subsection{Message Dissemination Schemes}
\label{sec::messageDisseminationSchemes}
Message dissemination schemes are algorithms that specify how a forwarding nodes is selected in a VANET. The selection of a forwarding node is performed based on some criteria such as distance between senders and receivers, number of received messages by a node, probabilities associated with nodes, topology of the network, etc~\cite{DBLP:journals/mis/SanguesaFGMCC16}. In this paper, two schemes --a distance-based and a counting-based scheme-- are modeled using the proposed framework. 

The distance-based scheme, called TLO (The Last One) \cite{TLO}, makes use of location information of the vehicles to select the forwarding node. In this scheme, upon a message broadcast, the farthest receiver in the range of the sender is selected as the forwarding TLO node. Other vehicles in the range know that they are not the farthest node and do not forward the received message. However, they wait for a while to make sure of successful broadcast of the TLO node. Receiving the warning message from the TLO node, means that the sending of the message has been successful and they do not forward the warning message. Otherwise, the algorithm is run once again to select the next TLO forwarding node.

In the counting-based scheme \cite{counting-based-scheme}, an integer number is defined as counter threshold. Each receiving node counts the number of received messages in a time interval. At the end of that time interval, the receiver decides on being a forwarding node based on the comparison of the value of its counter and the value of counter threshold. If the value of the counter is greater than the value of counter threshold, the receiver assumes that enough warning messages are disseminated in its vicinity; therefore, it avoids forwarding the message. Otherwise, the receiver forwards the warning message.

\subsection{Analysis Techniques}
Different analysis techniques have been developed for the analysis of message dissemination schemes in VANETs. Simulation-based approaches are widely used 
for the analysis of applications of in this domain. Gama et. al. developed a model and analyzed three different message dissemination schemes using Veins simulator \cite{DBLP:conf/iwcmc/GamaNCSMD17}. Sanguese et. al. have used ns-2 simulator in two independent works regarding the selection of optimal message dissemination scheme. In \cite{DBLP:conf/mswim/SanguesaFGMCCM13}, they aim to select the optimal broadcasting scheme for the model in each scenario and in \cite{DBLP:journals/comcom/SanguesaFGMCCM15}, the selection of the optimal scheme is performed for each vehicle based on vehicular density and the topological characteristics of the environment where the vehicle is located in. In a more comprehensive work \cite{DBLP:journals/mis/SanguesaFGMCC16} authors have developed a framework in ns-3 simulator for comparing different schemes. Note that although this approach is used in many applications, it does not guarantee correctness of results as it does not consider concurrent execution of system components.

%%%%%%
Another technique used for the analysis of WMD in VANETs is the analytical approach. In this approach, a system is modeled by mathematical equations and the analysis is performed by finding solutions to the equation system. For example, in \cite{DBLP:journals/tits/SaeedMPPL19}, Saeed et. al. have derived difference equations that their solutions yield the probability of all vehicles receiving the emergency warning message. This value is computed as a function of the number of neighbors of each vehicle, the rebroadcast probability, and the dissemination distance. In another work, a probabilistic multi-hop broadcast scheme is mathematically formulated and the packet reception probability is reported for different configurations, taking into account the topology of the network and as a result, major network characteristics such as vehicle density and the number of one-hop neighbors \cite{DBLP:conf/wd/GholibeigiH16}. This approach guarantees achieving correct results but it is not modular and developing mathematical formula needs a high degree of user interaction and a high degree of expertise.

As the third technique, model checking is a general verification approach which provides ease of modeling the same as simulation-based approaches in addition to guaranteeing the correctness of results due to its mathematical foundation. To the best of our knowledge, there is no framework which provides model checking facilities for the analysis of WMD schemes in VANETs.

\section{Rebeca Language}
\label{sec:rebeca}

Rebeca is a modeling language based on Hewitt and Agha's actors \cite{Agha1987}. Actors in Rebeca are independent units of concurrently running programs that communicate with each other through message passing. The message passing is an asynchronous non-blocking call to the actor's corresponding message server. Message servers are methods of the actor that specify the reaction of the actor to its corresponding received message. In the Java-like syntax of Rebeca, actors are instantiated from reactive class definitions that are similar to the concept of classes in Java. Actors in this sense can be assumed as objects in Java. Each reactive class declares the size of its message buffer \footnote{Message queue in Rebeca and message bag in Timed Rebeca}, a set of state variables, and the messages to which it can respond. Reactive classes have constructors with the same name as their reactive class, that are responsible for initializing the actor’s state.

Basically, in Rebeca the concept of known rebecs was introduced for an actor to specify the actors to which it can send message. However, to implement applications in ad-hoc networks, a more flexible sending mechanism is needed. Two Rebeca extensions b-Rebeca \cite{b-rebeca} and w-Rebeca \cite{w-rebeca} have been proposed to provide more complex sending mechanism. In b-Rebeca the concept of known rebecs is eliminated and it is assumed that the only communication mechanism among actors is broadcasting; hence, only a fully connected network can be modeled. Note that the type of broadcasting introduced in b-Rebeca is not the same as the location-based broadcasting in VANETs. In location-based broadcasting, only the actors in the range of each other are connected in the Rebeca model. Regarding this assumption, a counter-based reduction technique is used in b-Rebeca to reduce the state space size of the model making it impossible to send messages to a subset of actors.

The other extension w-Rebeca, which is developed for model checking of wireless ad-hoc networks, uses an adjacency matrix in the model checking engine, to consider connectivity of actors. In this approach, by random changes in the value of adjacency matrix, all the possible topologies of the network are considered in the model checking. Note that users are allowed to define a set of topological constraints and the topologies that do not fulfill the constraints are not considered in the model checking. w-Rebeca does not support timing in the model which is essential for developing models in the domain of VANET; since there are some real-time properties that need to be considered. Besides, considering all possible topologies --some of which may not be possible in the reality of the model-- results in a bigger state space for the model. In addition, considering these infeasible topologies, may cause false-negative results when checking correctness properties.

\subsection{Explaining Rebeca by the Example of Counting-Based Scheme}
\label{sec:Counting-model}

In this section, we introduce Timed Rebeca \cite{timed-rebeca} using the example of the counting-based scheme introduced in the previous section. A Timed Rebeca model consists of a number of reactive class definitions which provide type and behavior specification for the actors instantiated from them. There are two reactive classes \texttt{BroadcastingActor} and \texttt{Vehicle} in the implementation of counting-based WMD in \FeriVANca as shown in Listing~\ref{code::counting-based}.
\begin{lstlisting}[language=rebeca, caption=Counting-based scheme in Timed Rebeca, label=code::counting-based]
env int RANGE = 10;
env int THRESHOLD_WAITING = 4;
env int MESSAGE_SEND_TIME = 1;
env int C_THRESHOLD = 3;
abstract reactiveclass BroadcastingActor (5) {
	statevars {	int id, x, y; }	
	abstract msgsrv receive(int data);
	void broadcast(int data) { ... }
	double distance(BroadcastingActor bActor, BroadcastingActor cActor){...}
}
reactiveclass Vehicle extends BroadcastingActor(5){
	statevars{
		boolean isAV;
		int direction, latency, destX, destY, counter;
	}
	Vehicle (/*List of Parameters*/){	
		/*Variables Initializations*/
		if (isAV) {
			 self.alertAccident();
		} else
			self.move() after(latency);
	}
	msgsrv alertAccident(){ ...	}
	msgsrv move() { ... }	
	msgsrv stop () { ... }
	msgsrv finishWait(int hop) { ... }
	msgsrv receive(int hopNum) { ... }
}
main {
	Vehicle v1():(0,0,10,RIGHT,1,10,10,true), v2():(1,10,0,UP,2,10,10,false), 
		v3():(2,-1,0,RIGHT,1,10,0,false), v4():(3,0,1,DOWN,2,0,-10,false), 
		v5():(4,3,0,LEFT,1,-10,0,false); 
}
\end{lstlisting}

Each reactive class consists of a set of state variables and a message bag with the size specified in parentheses after reactive class name in the declaration. For example, reactive class \texttt{Vehicle} has state variables  \texttt{isAv}, \texttt{direction}, \texttt{latency}, \texttt{counter}, etc. The size of the message bag for this reactive class is set to five. The local state of each actor consists of the values of its state variables and the contents of its message bag. Being an actor-based language, Timed Rebeca benefits from asynchronous message passing among actors. Upon receiving a message, the message is added to the actor's message bag. Whenever the actor takes a message from the message bag, the routine which is associated with that message is executed. These routines are called message servers and are implemented in the body of reactive classes.

As depicted in Listing~\ref{code::counting-based}, the message servers of the reactive class \texttt{Vehicle} are \texttt{move}, \texttt{receive}, \texttt{alertAccident}, \texttt{stop}, and \texttt{finishWait}. In order for an actor to be able to send a message to another actor, the sender has to have a direct reference to the receiver actor. 
For example, in Line 19, the message \texttt{alertAccident} is sent to \texttt{self} which represents a reference to the actor itself.
%%%%% message sending aaddi ro tozi bede ba mohtavaye constr. vehicle. vagar na khar-id bi khar-id -> anjaam naddad vali raftim kharid! -------- hal shode o men bab e yadegari moonde
However, in order to model a WMD scheme in VANETs, the warning message should reach actors which are in the range of the sender actor. In other words, actors should receive messages based on some criteria, i.e., their location in this application. We used inheritance mechanism of Timed Rebeca to implement this customized sending strategy.

\subsection{Customized Message Sending in \FeriVANca} 

In object-oriented design, inheritance mechanism enables classes to be derived from another class and form a hierarchy of classes that share a set of attributes and methods. Using this approach, we encapsulated broadcasting mechanism in a reactive class called \texttt{BroadcastingActor} and all other behaviors of vehicles are implemented in \texttt{Vehicle} reactive class which is derived from \texttt{BroadcastingActor}. In \texttt{BroadcastingActor}, the \texttt{broadcast} method shown in Listing~\ref{code::broadcast} mimics the sending mechanism of vehicles in VANET.

 As mentioned before, broadcasting data, results in receiving a message containing that data by the vehicles in the range of the sender actor. In the body of this method, all actors --that are derived from \texttt{BroadcastingActor}-- are examined in terms of their distance to the sender (Line 5). If the distance between an actor and the sender is less than the specified threshold, called \texttt{RANGE} (Line  6), the data is sent to the actor by asynchronous message server call of \texttt{receive} (Line 7). As \texttt{BroadcastingActor} has no idea about the behavior of vehicles, upon receiving the \texttt{receive} message, the template method design pattern \cite{Gamma:1995:DPE:186897} is used in the implementation of \texttt{receive}. So, the \texttt{receive} message server is defined as an abstract message server in \texttt{BroadcastingActor} and its body is implemented in \texttt{Vehicle}. The behavior of the WMD scheme is implemented in \texttt{Vehicle}.

\begin{lstlisting}[language=rebeca, caption=Body of broadcast Method in Broadcasting Actor, label=code::broadcast ]
void broadcast(int data) {
	ArrayList<ReactiveClass> allActors = getAllActors();
	for(int i = 0; i < allActors.size(); i++) {
		BroadcastingActor ba = (BroadcastingActor)allActors.get(i);
		double distance = distance (ba , self);
		if(distance < RANGE) {
			ba.receive(data) after (MESSAGE_SEND_TIME);
		}
	}
}
double distance(BroadcastingActor bActor , BroadcastingActor cActor){
	return sqrt(pow(cActor.x - bActor.x, 2) + pow(cActor.y - bActor.y, 2));
}
\end{lstlisting}

\subsection{Counting-Based Scheme in \FeriVANca}
 
For the case of counting-based scheme, three message servers \texttt{alertAccident}, \texttt{finishWait}, and \texttt{receive} provide the behavior of the scheme. When \texttt{Vehicle} actors are instantiated, their constructor methods are executed resulting in sending one of the following messages to themselves: 
\begin{itemize}
\item
\texttt{alertAccindent}: sent by the accident vehicle to start the WMD algorithm (Line 8)
\item
\texttt{move}: sent by the other actors to begin moving with their pre-defined \texttt{latency}; an actor performs this through sending \texttt{move} message periodically to itself (Line 10).

\end {itemize}
The algorithm of counting-based scheme begins by serving \texttt{alertAccident} message in the accident vehicle. Upon the execution of \texttt{receive}, if the \texttt{counter}, which is initially set to zero for all actors (Line 6), is zero -- meaning that it is the first time the actor is receiving the warning message-- a watchdog timer is started. This is implemented by sending the \texttt{finishWait} message to the actor itself with the arrival time of  \texttt{THRESHOLD\_WAITING}. In addition, the value of \texttt{counter} is set to one to indicate that this is the first call of  \texttt{receive} (Lines 20-22). The next calls of  \texttt{receive} result in increasing the value of  \texttt{counter}, as the representer of the number of received warning messages. When message server \texttt{finishWait} is executed by an actor, showing that the watchdog timer is expired, the value of \texttt{counter} is compared with the threshold considered for the counter (\texttt{C\_THRESHOLD}). By not exceeding the threshold, i.e. the area around the actor is not covered by enough number of warning messages, the actor broadcasts the warning message (Lines 16 and 17).

\begin{lstlisting}[language=rebeca, caption=Body of message servers in Vehicle Actor, label=code::vehicleMethods]
reactiveclass Vehicle extends BroadcastingActor(5){

	statevars{ ... }
	Vehicle (...){
		...
		counter = 0;
		if (isAV) {
			 self.alertAccident();
		} else
			self.move() after(latency);
	}
	msgsrv alertAccident(){
		broadcast(0);
	}
	msgsrv finishWait(int hop){
		if (counter < C_THRESHOLD)
			broadcast(hop++);	
	}
	msgsrv receive(int hopNum) {
		if (counter == 0) {
			finishWait(hopNum) after (THRESHOLD_WAITING);
			counter = 1;
		} else {
			counter++;	
		}
	}
}
\end{lstlisting}

\subsection{Reusability of \FeriVANca}

To illustrate the reusability of \FeriVANca, we show how the model of the counting-based scheme can be altered to present another scheme (the TLO scheme) by making minor modifications to the code. At the first step, we implemented the algorithm in a method called \texttt{runTLO}. As shown in Listing~\ref{code::TLOCode} body of the message servers \texttt{finishWait} and \texttt{receive} are rewritten to mimic the behavior of the scheme in the event of expiration of the watchdog timer and receiving a warning message respectively.

As explained in Section~\ref{sec::messageDisseminationSchemes}, in the TLO scheme, upon receiving the warning message for the first time, the \texttt{runTLO} method is called. In the body of this method, if the value of state variable \texttt{received} is false --meaning that the actor has not received the duplicate warning message from a selected TLO node as a sign of its successful broadcast--, the \texttt{isTLO} method is called. This method is implemented in the \texttt{BroadcastingActor} and checks if the actor is the furthest node in the range of the sender and returns the result as a boolean value. If the return value is true, the actor is the last one in the range and is selected as the TLO node to forward the warning message; so, it broadcasts the message by increasing the value of \texttt{hupNum} by one (Line 15). Then the value of \texttt{received} is set to true to show that broadcasting has been successful. In case the actor is not the last one in the range (Line 17), the actor should wait for a while to make sure that the selected TLO node has successfully broadcasted the warning message. To this end, the actor sets the value of \texttt{isWaiting} to true to show that the actor is in the waiting mode, and then sets the watchdog timer by sending message \texttt{finishWait} to itself by execution time of \texttt{THRESHOLD\_WAITING} (Line 19).
The message server \texttt{receive}, like in the previous scheme, mimics receiving the warning message. In the body of this message server, the value of the state variable \texttt{isAware} is set to true to show that the actor is aware of the accident due to the receiving of the warning message. Then, if the value of \texttt{isWaiting} is false, meaning that the actor is not in the waiting mode, \texttt{isTLO} is executed to select the TLO forwarding node. Otherwise, \texttt{isWaiting} is set to false since this message is interpreted as a successful broadcast of the TLO node. The \texttt{finishWait} message server is executed upon expiration of the watchdog timer and it checks the value of \texttt{isWaiting}. In the case of false value for \texttt{finishWait}, the actor has not received the warning message from the selected TLO node, so, \texttt{runTLO} is called to select the next TLO forwarding node.

\begin{lstlisting}[language=rebeca, caption=Needed modifications for TLO scheme, label=code::TLOCode, multicols=2]
msgsrv finishWait(int hopNum) {
	if (isWaiting)
		runTLO(hopNum);
}	
msgsrv receive(int hopNum) {
	isAware = true;
	if(!isWaiting)
		runTLO(hopNum);
	else
		isWaiting = false;
}
void runTLO(int hopNum) {
	if (!received) {
		if (isTLO()) {
			broadcast(hopNum++);
			received = true;
		} else {
			isWaiting = true;
			self.finishWait(hopNum) after(THRESHOLD_WAITING);
		}	
	}
}
\end{lstlisting}

\section{Experimental Results}
\label{sec:exprRes}

To demonstrate the applicability of \FeriVANca, both of the schemes presented in the former section are analyzed in different configurations. As mentioned before, concurrent behaviors of the system components may cause uncertainty which are clearly observable in the presented scenarios, but may not be detected using simulation-based techniques. For the case of the TLO scheme, we showed that nondeterminism causes starvation and for the case of the counting-based scheme, it causes different results in the performance of the algorithm. Furthermore, we made clear that the approach is scalable regarding the number of cars with traffic patterns that do not contain congested areas. Note that the following experiments have been executed on a Macbook Air with Intel Core i5 1.3 GHz CPU and
8GB of RAM, running macOS Mojave 10.14.2 as the operating system. Development of these experiments are performed in Afra, modeling and verification IDE of Rebeca family languages \cite{DBLP:journals/csur/BoerSHHRDJSKFY17}.

\subsection{Starvation Scenario in TLO Scheme}
\label{sec:starvation}
\begin{figure}
\centering
\subfigure[Accident between A and B]{
\label{fig::TLOSen12}
  \centering
  \small{
   \includegraphics[width=.40\textwidth]{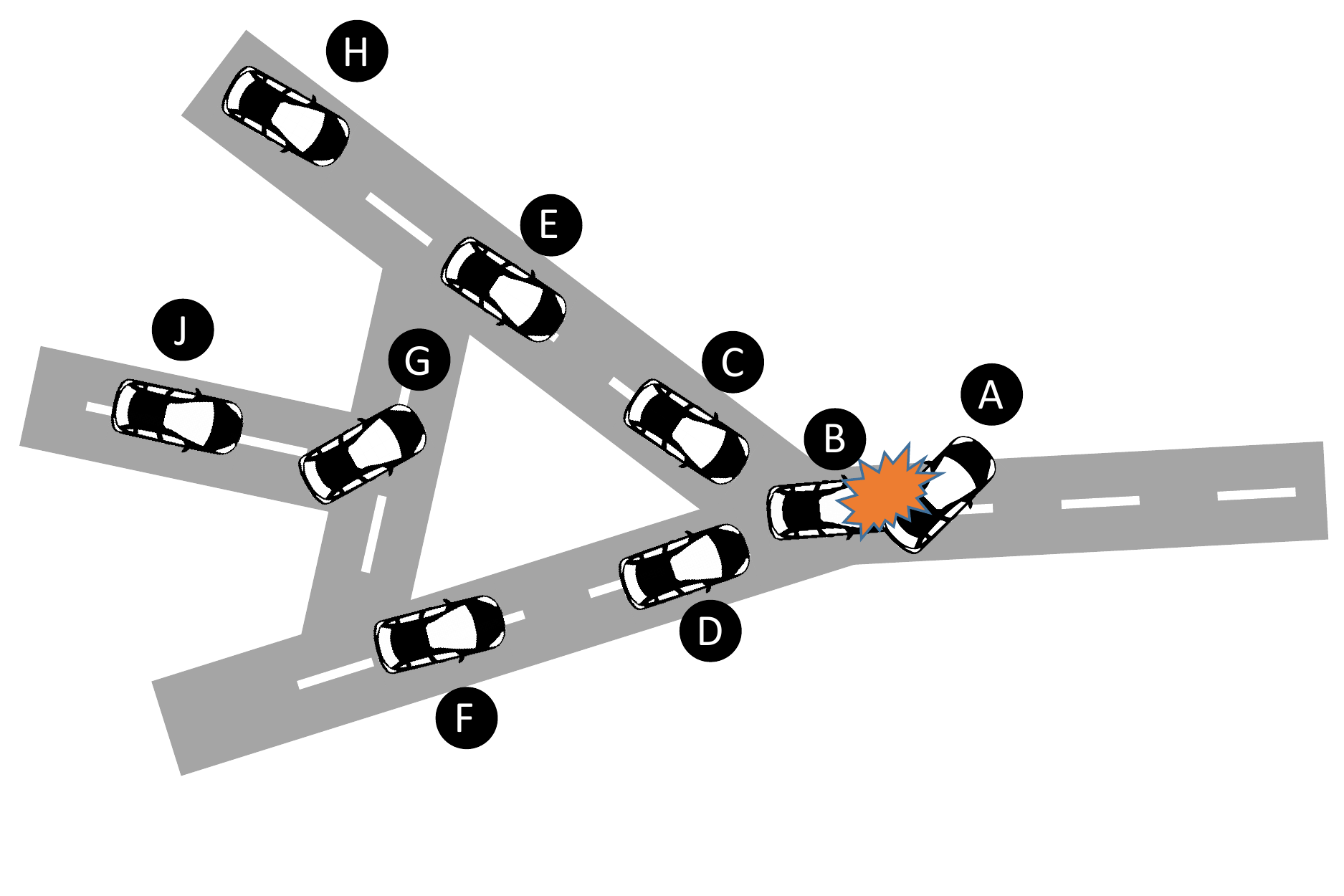}
  }
}
\qquad
\subfigure[B broadcasts the warning message]{
\label{fig::TLOSen22}
  \centering
  \small{
   \includegraphics[width=.40\textwidth]{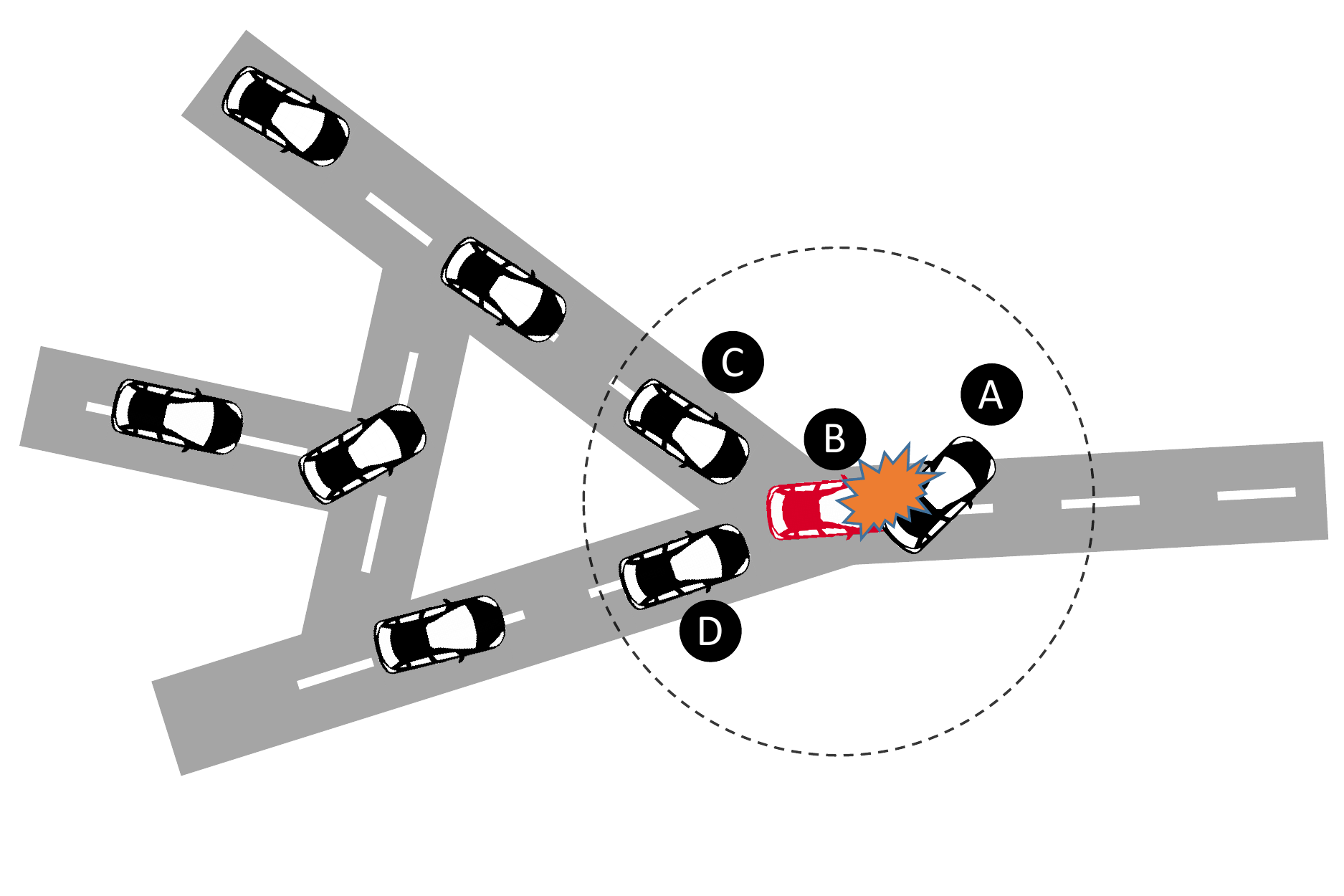}
  }
}
\subfigure[C and D are both selected as TLO nodes to forward the warning message]{
\label{fig::TLOSen32}
  \centering
  \small{
   \includegraphics[width=.40\textwidth]{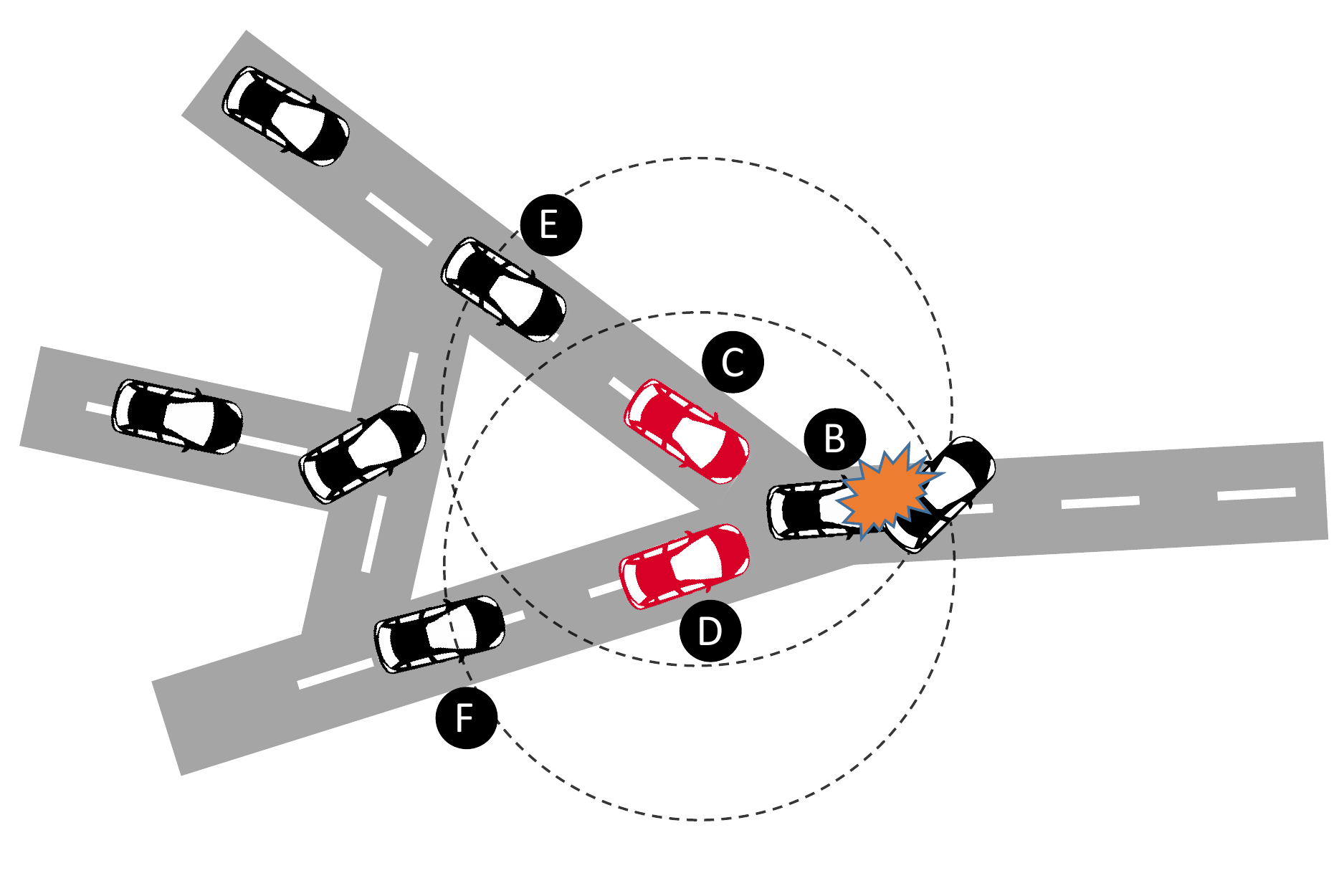}
  }
}
\qquad
\subfigure[Order of broadcasting between E and F results in two cases]{
\label{fig::TLOSen42}
  \centering
  \small{
   \includegraphics[width=.40\textwidth]{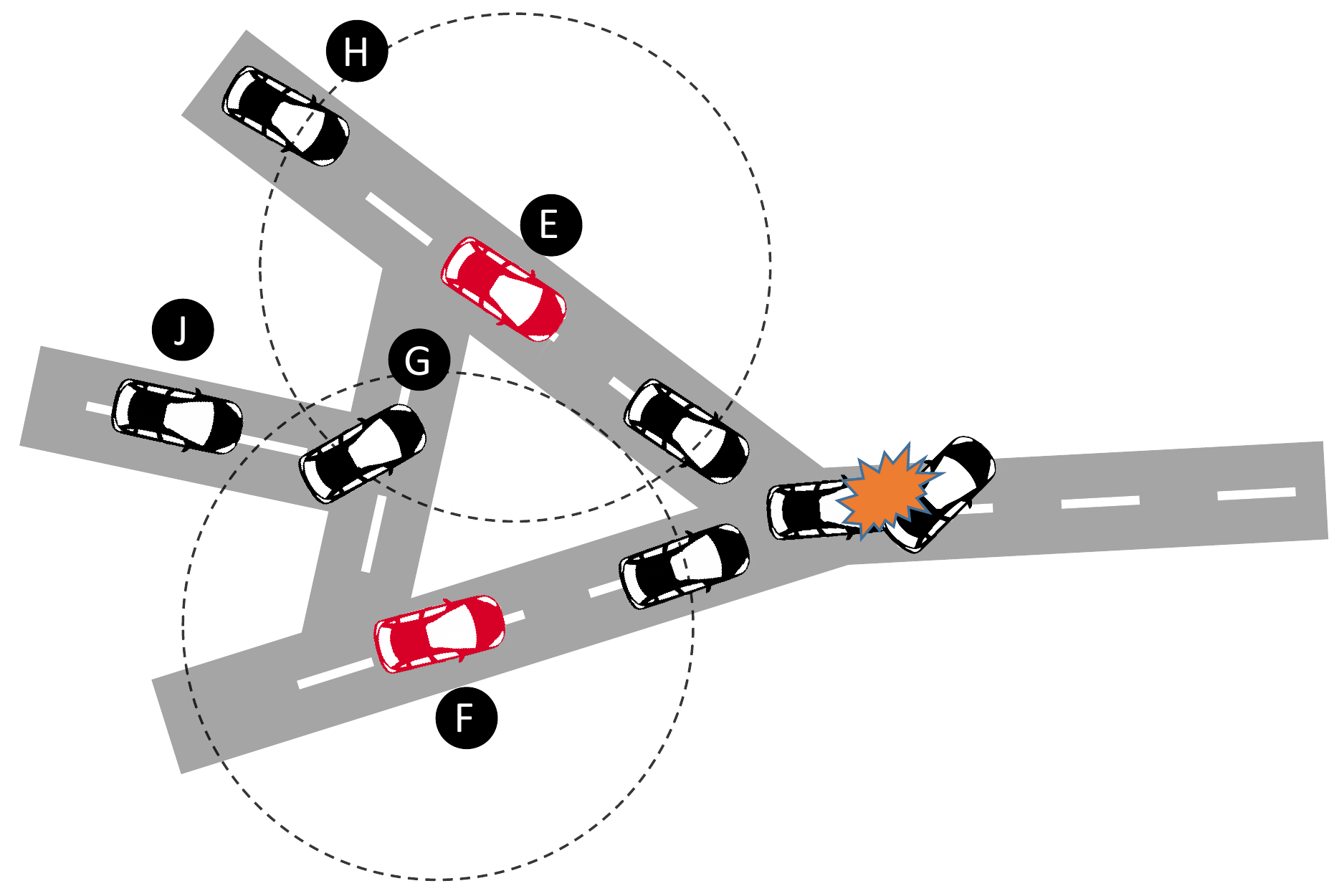}
  }
}
\caption{A scenario of TLO scheme which results in two execution alternatives that one of them causes starvation for vehicle J}
\label{fig::TLOSenario2}
\end{figure}

In this section, we present an observed scenario that using the TLO scheme  causes starvation and affects the reliability of the scheme in some executions. The steps of the scenario is depicted in Figure~\ref{fig::TLOSenario2}. In~\ref{fig::TLOSen12}, position of the vehicles is shown in the time of the accident between vehicles A and B. In the next step, vehicle B starts broadcasting the warning message and vehicles C and D receive the message as they are in the range of B (Figure~\ref{fig::TLOSen22}). Upon receiving the warning message, these vehicles execute the TLO algorithm and since they both have the same distance from B, they forward the received warning message and the vehicles E and F receive the warning message from these two vehicles. When vehicles E and F execute the TLO algorithm, racing between the following two scenarios happen.
\begin{enumerate}
\item 
\textbf{E broadcasts before F:} vehicles G and H receive the warning message from E. Upon execution of TLO algorithm by G and H, Vehicle H is selected as the TLO forwarding node and forwards the message. Meanwhile, vehicle G is waiting for receiving the warning message from H to make sure that the broadcasting has been successful. If in the waiting time of G, vehicle H forwards the warning message, the message will be interpreted as acknowledgement of the successful broadcast of H and although G is TLO node in this step, it will not forward the message. In this case, the vehicle J does not receive the warning message. 
\item 
\textbf{F broadcasts before E:} vehicle G receive the warning message from F and after the execution of TLO algorithm, it forwards the message as the selected TLO node and vehicle J will receive the warning message in this scenario.
\end{enumerate}

This example shows that concurrent execution of the algorithm in nodes causes nondeterministic behavior which may violate correctness properties of the application. To avoid such cases, all the possible nondeterministic behaviors have to be considered in any analysis framework. However, simulation-based techniques, commonly used for the analysis of these systems, fail to report a result by considering all the possible execution traces. This highlights the necessity of applying formal methods in the development of applications of VANETs with critical mission.

\subsection{Nondeterminism in Performance of the Counting-Based Scheme}
\label{nondet-performance-result}
\begin{figure}
\centering
\subfigure[Configuration of the scenario]{
\label{fig::nondet1}
  \centering
  \small{
   \includegraphics[width=.4\textwidth]{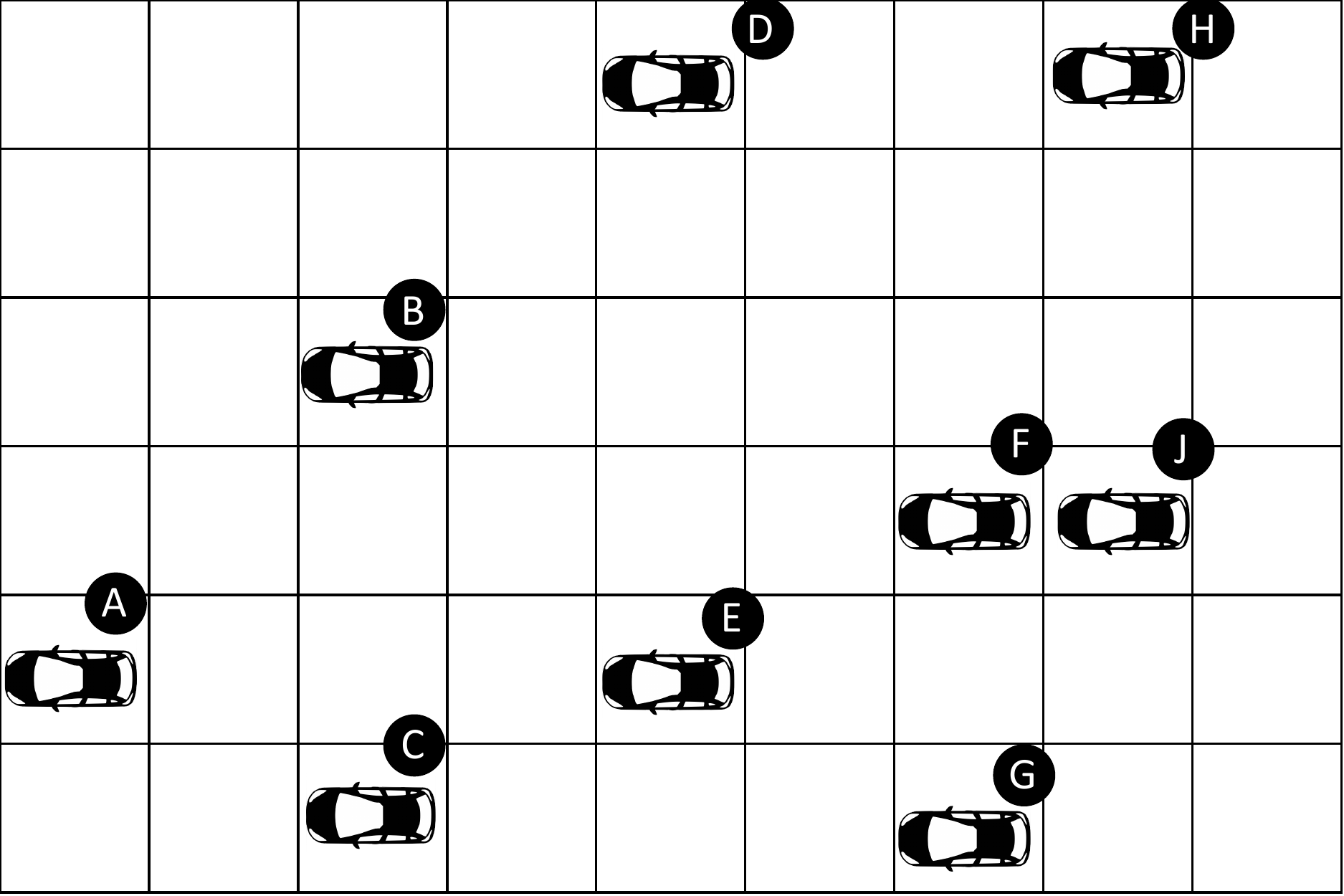}
  }
}
\qquad
\subfigure[Vehicle A starts broadcasting]{
\label{fig::nondet2}
  \centering
  \small{
   \includegraphics[width=.4\textwidth]{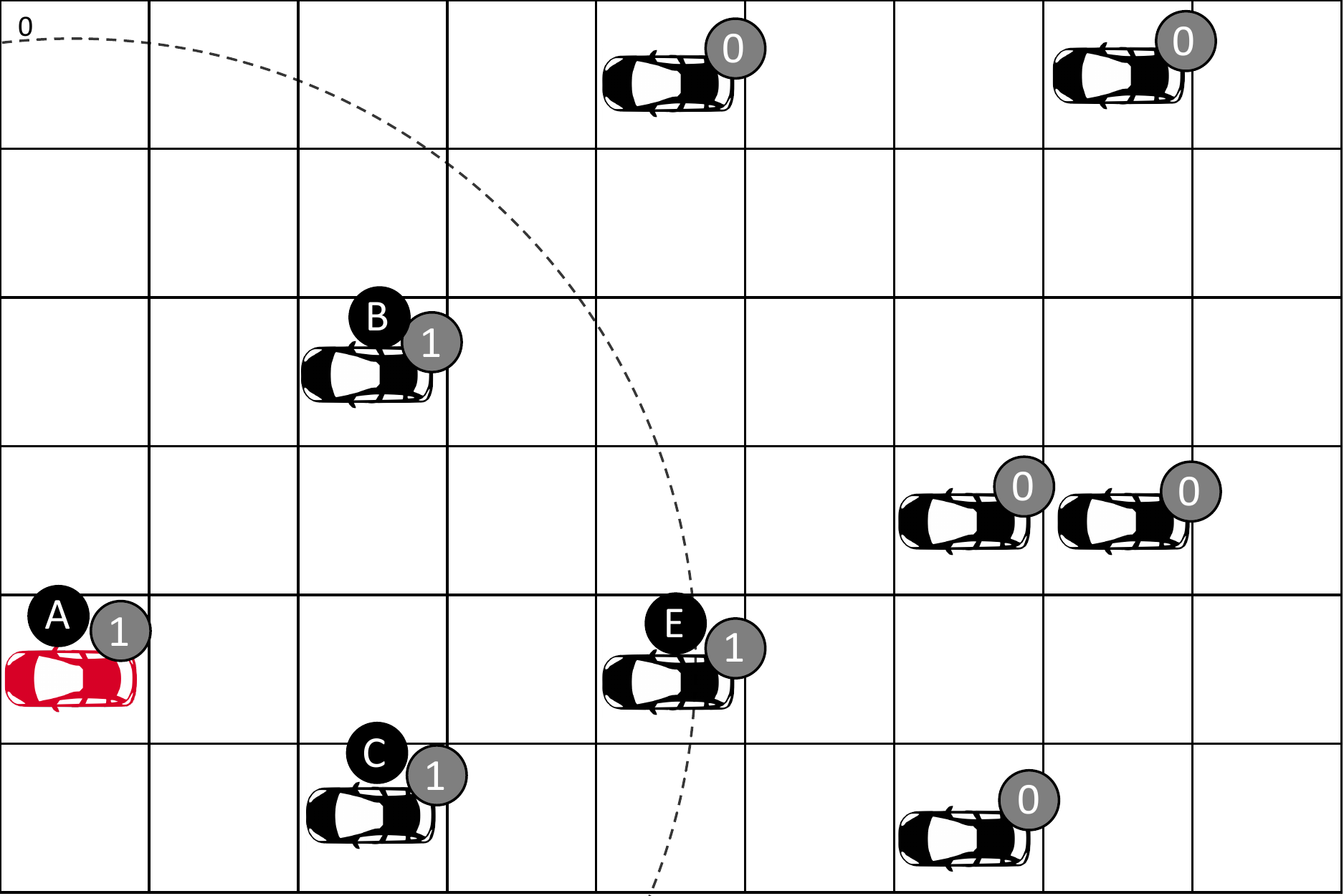}
  }
}
\subfigure[B broadcasts before expiration of the watchdog timer of E]{
\label{fig::nondet3}
  \centering
  \small{
   \includegraphics[width=.4\textwidth]{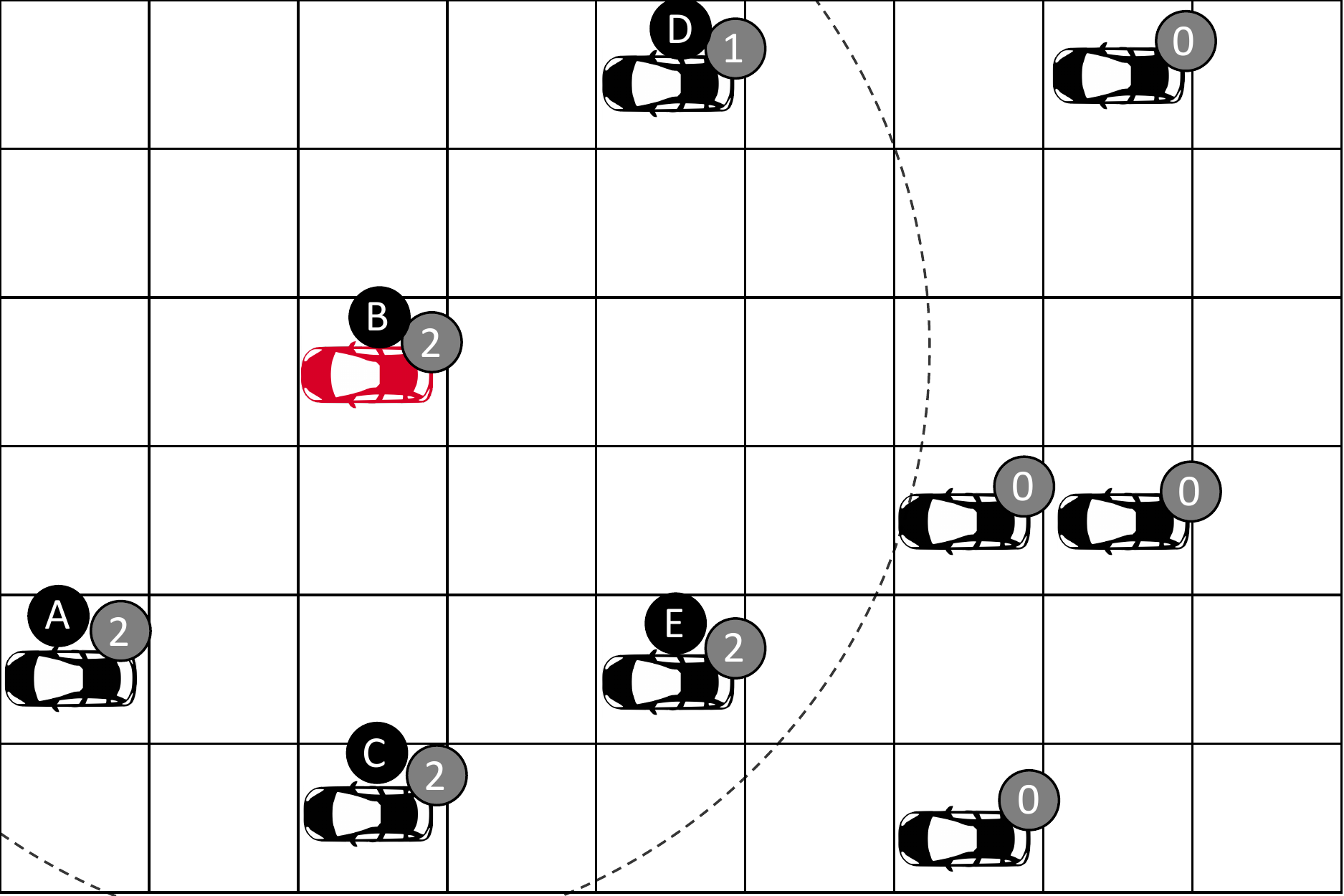}
  }
}
\qquad
\subfigure[D forwards the warning message]{
\label{fig::nondet4}
  \centering
  \small{
   \includegraphics[width=.4\textwidth]{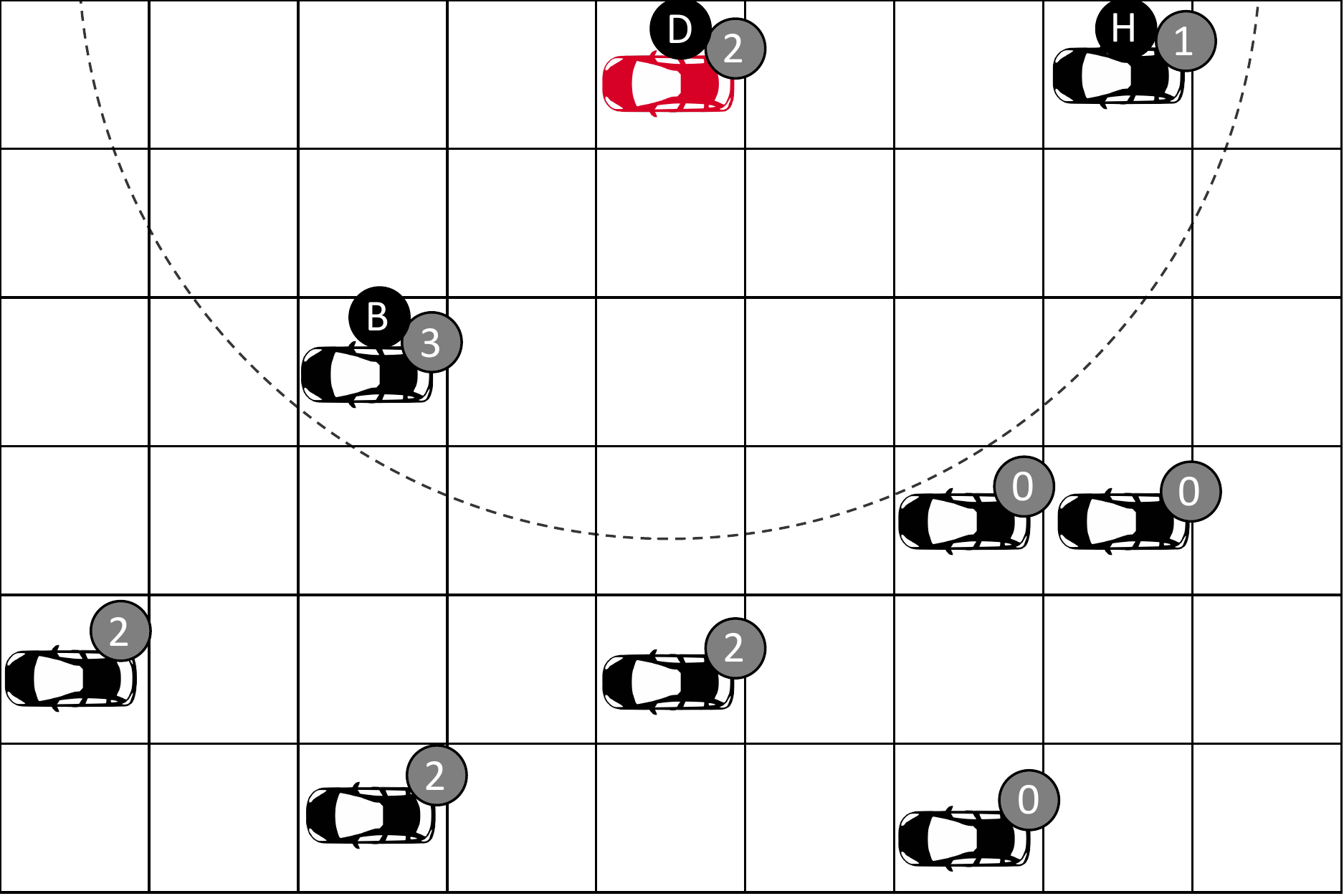}
  }
}
\subfigure[H rebroadcasts the message]{
\label{fig::nondet5}
  \centering
  \small{
   \includegraphics[width=.4\textwidth]{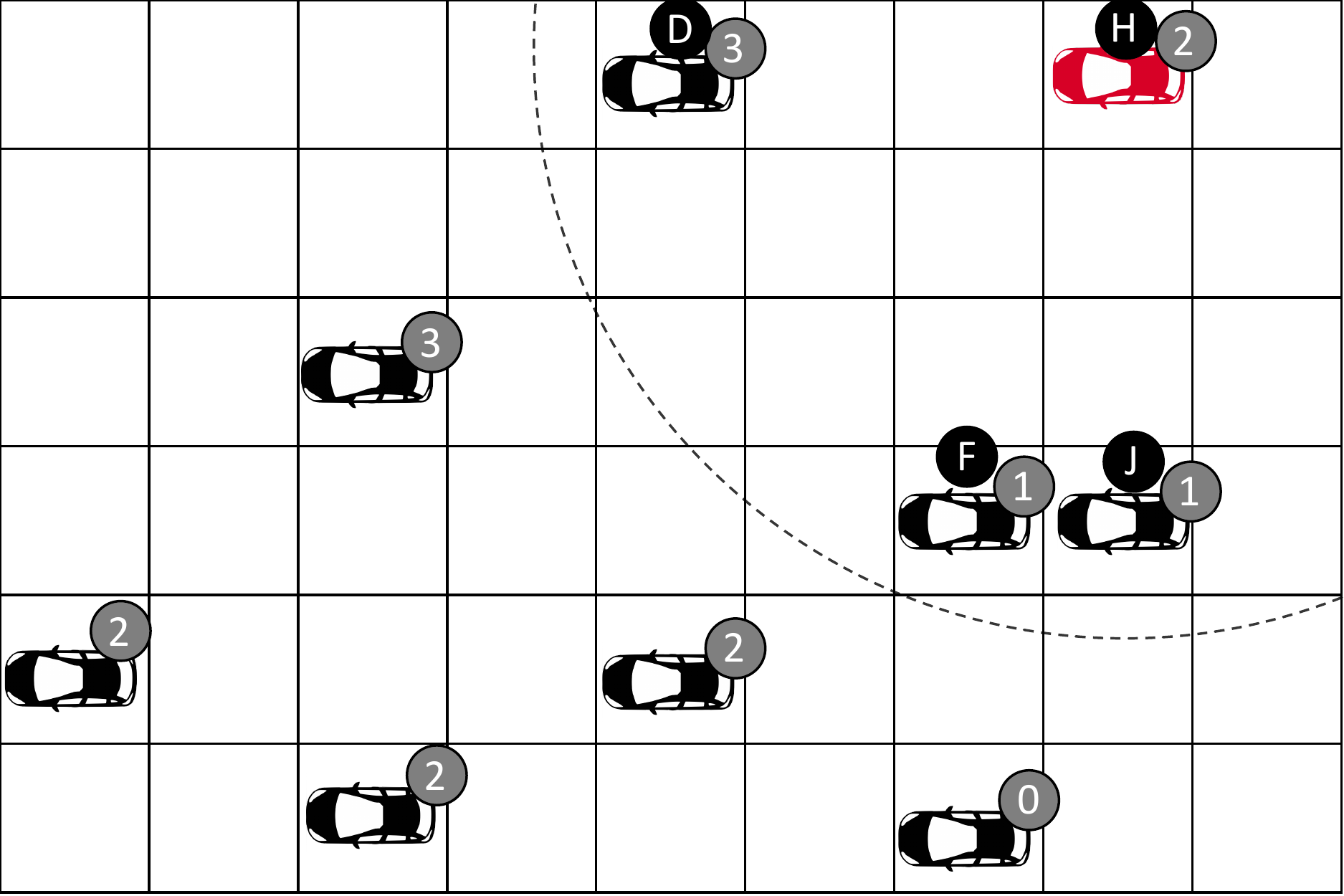}
  }
}
\qquad
\subfigure[F and (or) J forward(s) the message and algorithm finishes]{
\label{fig::nondet6}
  \centering
  \small{
   \includegraphics[width=.4\textwidth]{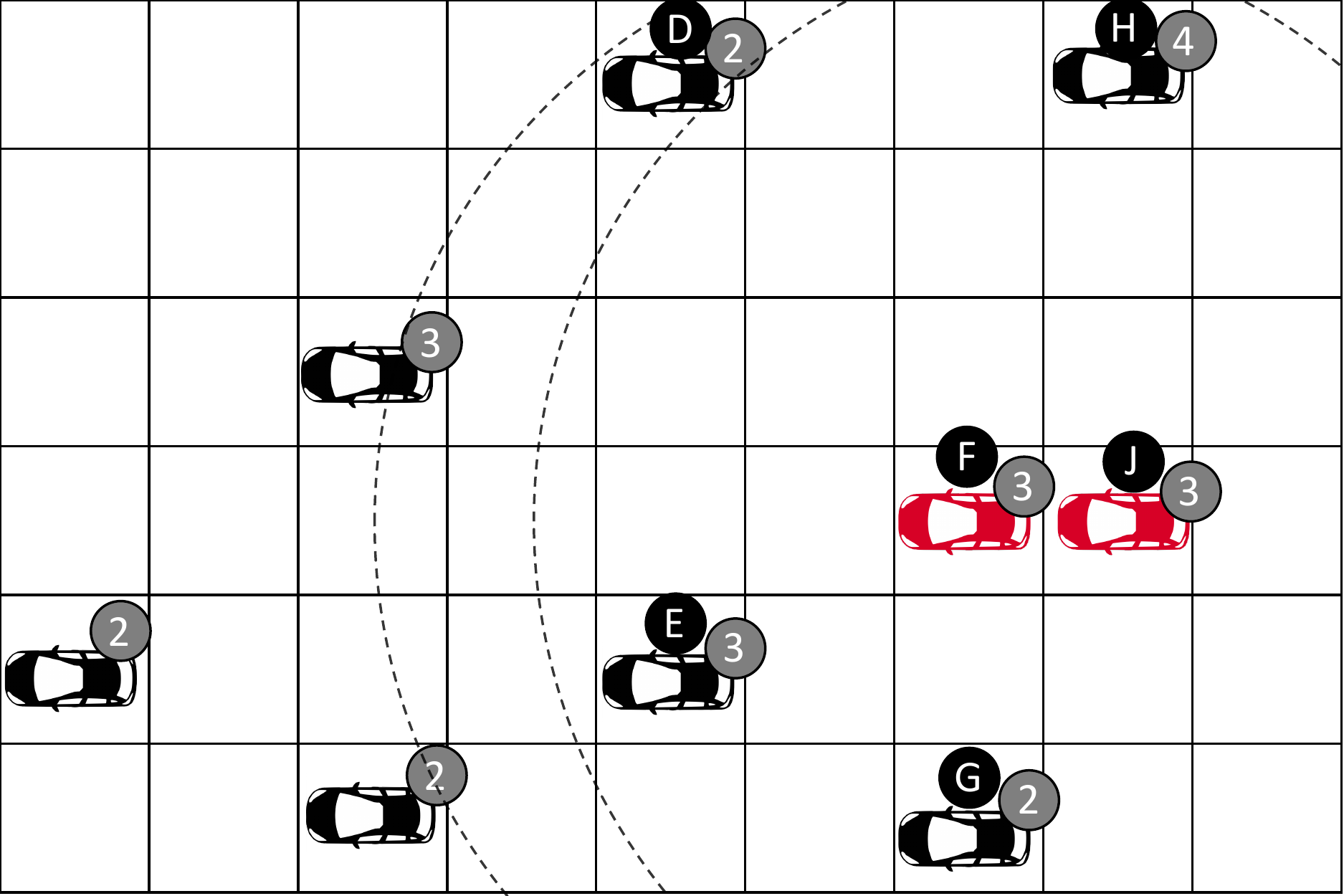}
  }
}

\caption{A case of the scenario for the counting-based scheme}
\label{fig::non-det}
\end{figure}

The configuration depicted in Figure~\ref{fig::nondet1} is used for the analysis of the counting-based scheme (explained in Section~\ref{sec:Counting-model}). In this scenario, the value of \texttt{C\_THRESHOLD} is set to 2 and the \texttt{RANGE} is set to 4. The scenario begins with the vehicle A broadcasting the warning message (Figure~\ref{fig::nondet2}). This broadcast results in increasing the counters of the vehicles A, B, C, and E by one. In the next round two following cases may happen.

\begin{figure}
\centering
\subfigure[E is selected as forwarder (instead of B as depicted in Figure~\ref{fig::nondet3})]{
\label{fig::nondet7}
  \centering
  \small{
   \includegraphics[width=.4\textwidth]{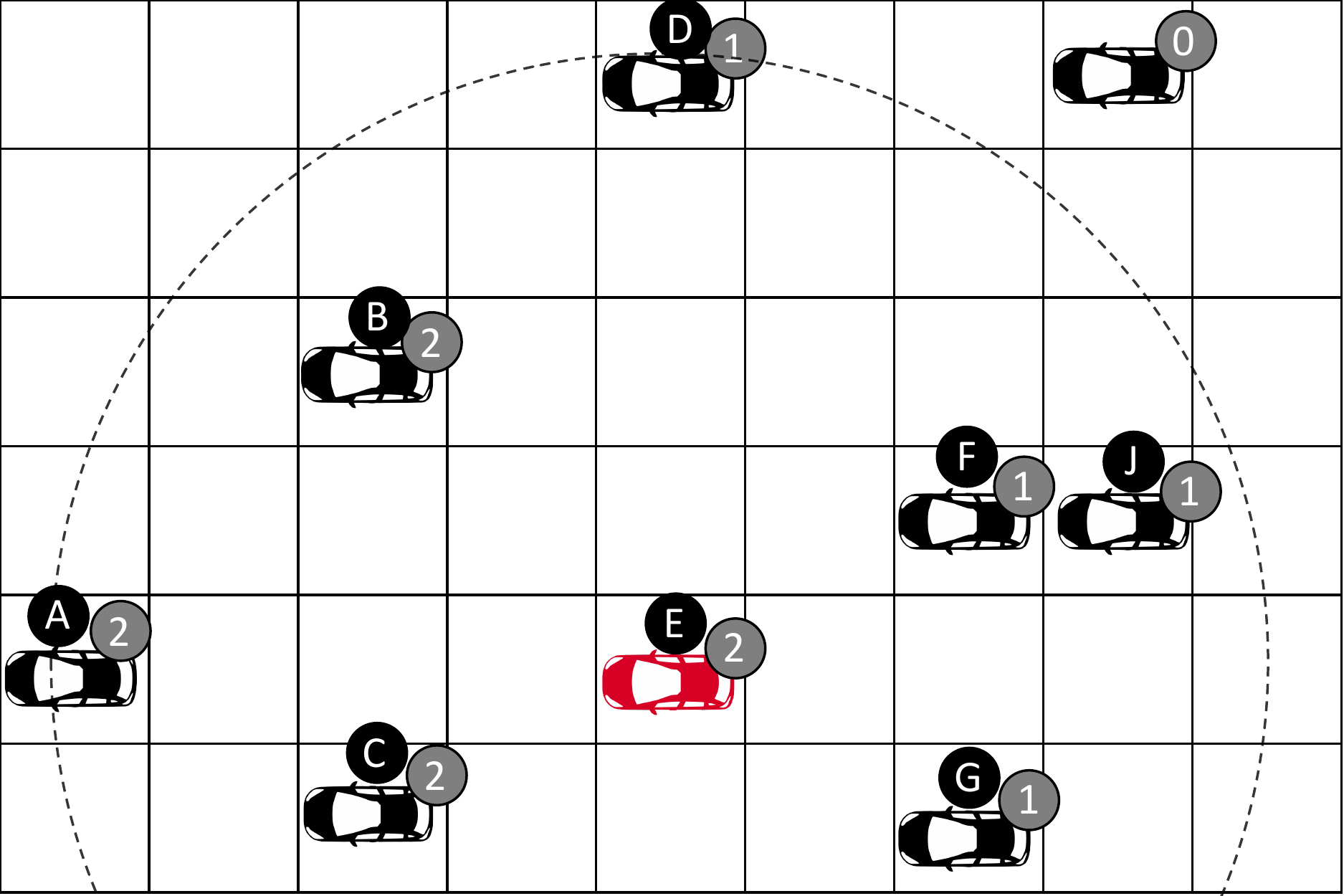}
  }
}
\qquad
\subfigure[F broadcasts the message and algorithm finishes]{
\label{fig::nondet8}
  \centering
  \small{
   \includegraphics[width=.4\textwidth]{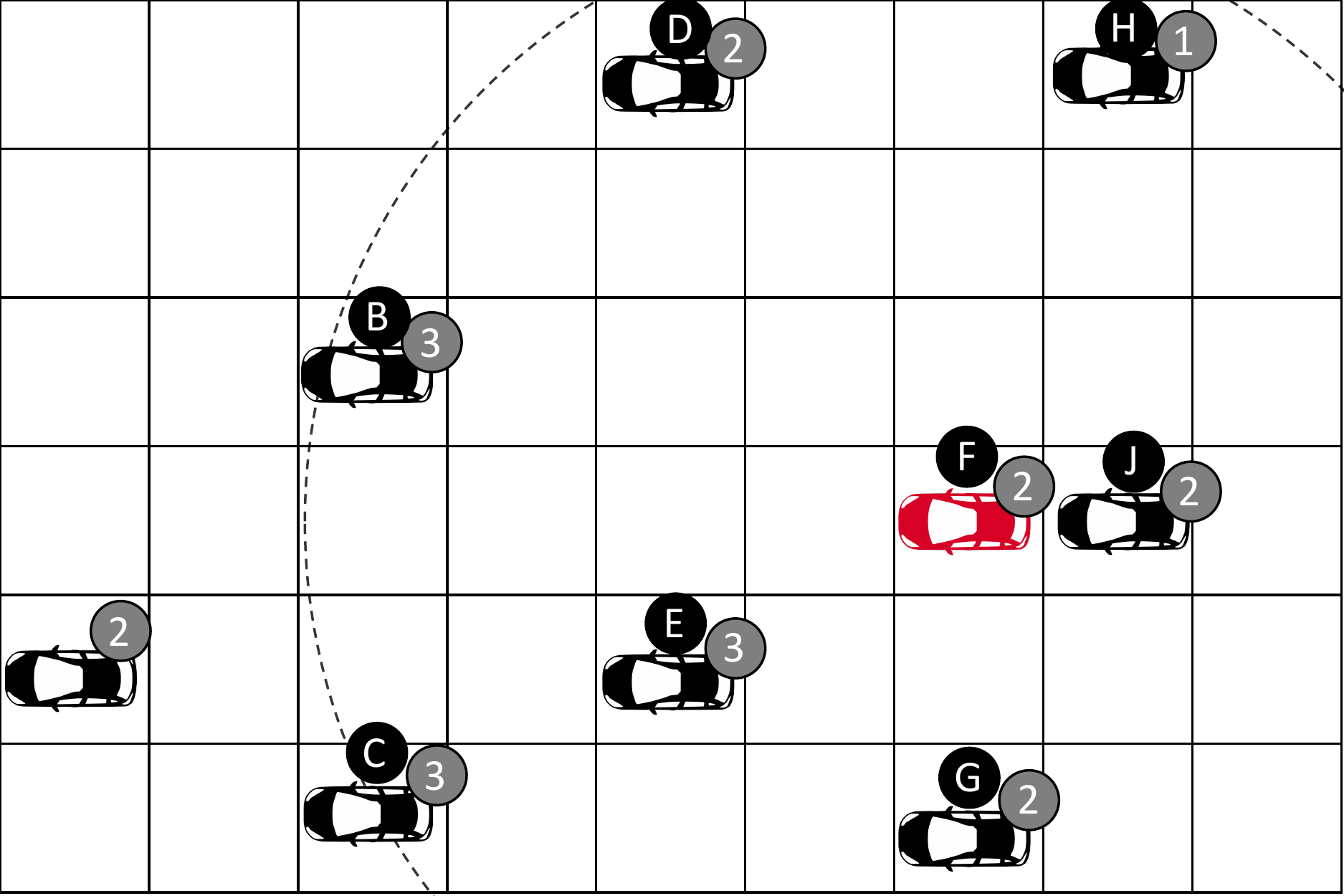}
  }
}
\caption{Another case of the scenario for the counting-based scheme}
\label{fig::non-det-s2}
\end{figure}

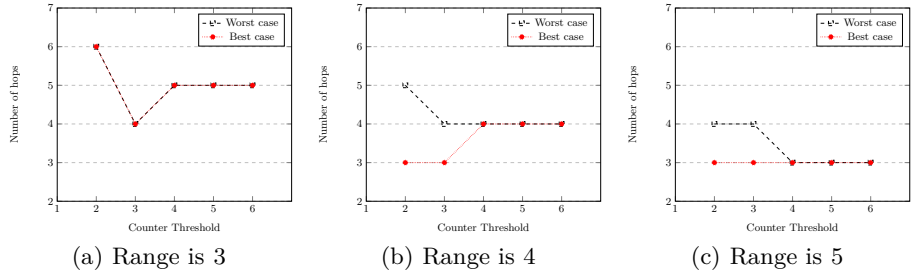
\begin{figure}  
\centering  
\subfigure[Range is 3]  
{  
 \begin{tikzpicture}[scale=.45]  
\begin{axis}[
    xlabel={Counter Threshold},
    ylabel={Number of hops},
    xmin=1, xmax=7,
    ymin=2, ymax=7,
    xtick={1,2,3,4,5,6},
    ytick={2,3,4,5,6,7},
    legend pos=north east,
    ymajorgrids=true,
    grid style=dashed,
]
 
\addplot[
    color=black,
    mark=square,dashed
    ]
    coordinates {
    (2,6) (3,4) (4,5) (5,5) (6,5)
    };
\addplot[
    color=red,
    mark=*,densely dotted
    ]
    coordinates {
    (2,6) (3,4) (4,5) (5,5) (6,5)
    };
    \legend{Worst case, Best case}
\end{axis}
 \end{tikzpicture}

}  
\subfigure[Range is 4]  
{  
\begin{tikzpicture}[scale=.45]  

\begin{axis}[
    xlabel={Counter Threshold},
    ylabel={Number of hops},
    xmin=1, xmax=7,
    ymin=2, ymax=7,
    xtick={1,2,3,4,5,6},
    ytick={2,3,4,5,6,7},
    legend pos=north east,
    ymajorgrids=true,
    grid style=dashed,
]
 
\addplot[
    color=black,
    mark=square,dashed
    ]
    coordinates {
    (2,5) (3,4) (4,4) (5,4) (6,4)
    };
\addplot[
    color=red,
    mark=*,densely dotted
    ]
    coordinates {
    (2,3) (3,3) (4,4) (5,4) (6,4)
    };
    \legend{Worst case, Best case}
\end{axis}
\end{tikzpicture}  

}  
% The only difference is here, where I have commented out an empty line.
\subfigure[Range is 5]  
{  

\begin{tikzpicture}[scale = 0.45]  

\begin{axis}[
    xlabel={Counter Threshold},
    ylabel={Number of hops},
    xmin=1, xmax=7,
    ymin=2, ymax=7,
    xtick={1,2,3,4,5,6},
    ytick={2,3,4,5,6,7},
    legend pos=north east,
    ymajorgrids=true,
    grid style=dashed,
]
 
\addplot[
    color=black,
    mark=square,dashed
    ]
    coordinates {
    (2,4) (3,4) (4,3) (5,3) (6,3)
    };
\addplot[
    color=red,
    mark=*,densely dotted
    ]
    coordinates {
    (2,3) (3,3) (4,3) (5,3) (6,3)
    };
    \legend{Worst case, Best case}
\end{axis}
\end{tikzpicture}  

}
\label{fig::plots}
\caption{Analysis results of the counting-based scheme with different values for the range and counter threshold (Note that Y axis shows the number of hops required for termination of the algorithm)}
\end{figure}

\begin{enumerate}
\item \textbf{The watchdog timer of vehicle E expires after receiving the message from B:} In this case, as the counter has reached the threshold, E does not forward the warning message as shown in Figure~\ref{fig::nondet3}. Following this case, the algorithm continues with vehicles D, H, and F being selected as forwarding nodes and rebroadcasting the message (Figures~\ref{fig::nondet4} to~\ref{fig::nondet6}. As a result, it takes 5 hops for all the vehicles to get informed of the warning message. Note that the same scenario happens when C forwards the message before the expiration of the watchdog timer of E.

\item \textbf{The watchdog timer of vehicle E expires before receiving warning message from B and C:} In this case, since the counter of E is less than the threshold, E must forward the warning message (Figure~\ref{fig::nondet7}). In the next step, vehicle F broadcasts the message and all non-informed vehicles receive the warning message and algorithm finishes in 3 hops.

\end{enumerate}
Achieving two different numbers for performance of this algorithm shows that beside correctness properties, providing guaranteed values for performance results requires applying formal verification techniques as well. We analyzed this scenario with different values for range and counter threshold, the result of three of them are shown in Figure~\ref{fig::plots}. The results show that this phenomenon is not rare and can be observed in many cases.

\subsection{Scalability Analysis}
\label{sec:scalability}
\begin{figure}
\centering
\subfigure[]{
\label{fig::max1}
  \centering
  \small{
   \includegraphics[width=\textwidth]{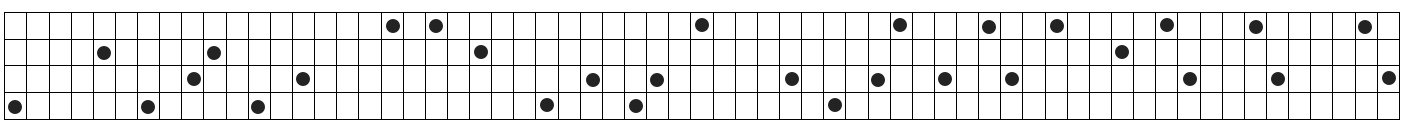}
  }
}
\subfigure[]{
\label{fig::max2}
  \centering
  \small{
   \includegraphics[width=\textwidth]{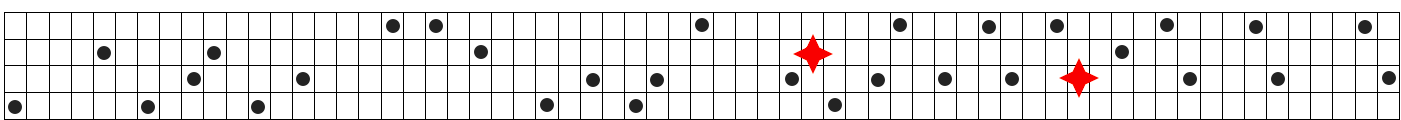}
  }
}
\caption{Configuration of the scenario used for scalability analysis}
\label{fig::scaleSenario}
\end{figure}

For the purpose of scalability analysis, we have modeled a four-lane street which contains about 30 vehicles. These vehicles are distributed in a way that there is no congested area in the street as shown in Figure~\ref{fig::max1}. The execution time of this model is 11 seconds and the number of reached states and transitions are 19,588 and 110,627 respectively. To determine the scalability, we added new cars in two ways. First, we increased the length of the street and added new vehicles to the tail of the street of Figure~\ref{fig::max1}. To avoid creating congested areas, we kept the same distribution while adding new vehicles. This way of scaling resulted in 15 seconds, 23,734 states, and 133,255 transitions for 35 vehicles and 18 seconds, 25,872 states, and 143,727 transitions for 40 vehicles (i.e. about 1.3 times more than the first case). As an estimation of the supported maximum size of the model regarding the state space size limit of Afra, the number of vehicles can be increased up to 100 if having distribution which does not create congested areas.
In the second way, new vehicles were added in a way to increase congestion in some areas (Figure~\ref{fig::max2}). Scaling in this way increases the execution time of the model to 120 seconds and the number of reached states and transitions to 157,086 and 1,265,839, respectively (i.e. about 10 times more than the previous case). This is because of the fact that in a congested area, the number of delivered warning messages to each vehicle grows rapidly and all the possible orders of execution for messages with the same execution time are considered in the model checking. This results in a sharp growth in the size of the state space and model checking time consumption.

\section{Conclusion and Future Work}
\label{sec:conclusion}

Lack of a framework for formal modeling and efficient verification of warning message dissemination schemes in VANETs is the main obstacle in using these schemes in real-world applications. In this paper, we presented \FeriVANca, an actor-based framework, developed using Timed Rebeca for modeling warning message dissemination schemes in VANETs. Model of schemes developed in \FeriVANca can be analyzed using Afra, the model checking tool of Timed Rebeca. We showed how warning message dissemination schemes can be modeled using \FeriVANca by implementing two of these schemes. Scenarios in these schemes were explored to illustrate the effectiveness of the approach in checking correctness properties and performance evaluation of the schemes. We further explained how easily the model of a scheme can be transformed to present another scheme by making minor modifications. Providing this level of guarantee in correctness and performance of warning message dissemination schemes, enables engineers to benefit from these schemes in the development of smart cars. 

Considering different members of Rebeca family modeling language, \FeriVANca can be used for addressing other characteristics of schemes such as their probabilistic behavior. Since Afra supports different members of Rebeca family, models with these characteristics can be analyzed using Afra.

\FeriVANca can be used for the analysis of scenarios with limited congested areas. However, to be able to use the framework for large-scale models containing congested areas, we are going to develop a partial order reduction technique. This reduction relies on the fact that reaction of a vehicle to received warning messages is independent of their sender; therefore, different orders of execution (interleaving) for messages received at the same time can be ignored without affecting the result of model checking.

\section*{Acknowledgments}
The work on this paper has been supported in part by the project ``Self-Adaptive Actors: SEADA'' (163205-051) of the Icelandic Research Fund and DPAC Project (Dependable Platforms for Autonomous Systems and Control) at M\"alardalen University, Sweden.

\renewcommand{\bibsection}{\section*{References}} 
% requried for natbib to have "References" printed and as section*, not chapter*
% Use natbib compatbile splncsnat style.
% It does provide all features of splncs03, but is developed in a clean way.
% Source: http://phaseportrait.blogspot.de/2011/02/natbib-compatible-bibtex-style-bst-file.html
\bibliographystyle{splncsnat}
\begingroup
  \ifluatex
    %try to activate if bibliography looks ugly
    %\sloppy
  \else
    \microtypecontext{expansion=sloppy}
  \fi
  \small % ensure correct font size for the bibliography
  \bibliography{paper}
\endgroup

% Enfore empty line after bibliography
\ \\
%
%All links were last followed on October 5, 2017.
\end{document}